# AutoFuse: Automatic Fusion Networks for Deformable Medical Image Registration

Mingyuan Meng [a, b], Michael Fulham [a, c], Dagan Feng [a], Lei Bi [a, b], and Jinman Kim [a, *]

[a] School of Computer Science, The University of Sydney, Australia.
[b] Institute of Translational Medicine, Shanghai Jiao Tong University, Shanghai, China.
[c] Department of Molecular Imaging, Royal Prince Alfred Hospital, Australia.
[*] Corresponding author: jinman.kim@sydney.edu.au

**Abstract —** Deformable image registration aims to find a dense non-linear spatial correspondence between a pair of images, which is a crucial step for many medical tasks such as tumor growth monitoring and population analysis. Recently, Deep Neural Networks (DNNs) have been widely recognized for their ability to perform fast end-to-end registration. However, DNN-based registration needs to explore the spatial information of each image and fuse this information to characterize spatial correspondence. This raises an essential question: what is the optimal fusion strategy to characterize spatial correspondence? Existing fusion strategies (e.g., early fusion, late fusion) were empirically designed to fuse information by manually defined prior knowledge, which inevitably constrains the registration performance within the limits of empirical designs. In this study, we depart from existing empirically-designed fusion strategies and develop a data-driven fusion strategy for deformable image registration. To achieve this, we propose an Automatic Fusion network (AutoFuse) that provides flexibility to fuse information at many potential locations within the network. A Fusion Gate (FG) module is also proposed to control how to fuse information at each potential network location based on training data. Our AutoFuse can automatically optimize its fusion strategy during training and can be generalized to both unsupervised registration (without any labels) and semi-supervised registration (with weak labels provided for partial training data). Extensive experiments on two well-benchmarked medical registration tasks (inter- and intra-patient registration) with eight public datasets show that our AutoFuse outperforms state-of-the-art unsupervised and semi-supervised registration methods.

## 1. Introduction

Image registration is a fundamental requirement for medical image analysis and has been an active research focus for decades [1][2]. Image registration spatially aligns medical images acquired from different patients, time-points, or scanners, which is a crucial step for a variety of medical tasks, such as tumor growth monitoring and population analysis [3], and also can be used for segmentation through registration-based label propagation [4]. Due to anatomy variations among patients or pathological changes such as tumor growth, medical images usually carry non-linear local deformations, especially for complex organs such as the cerebral cortex in the brain [5]. Therefore, different from the common natural image registration tasks, e.g., panorama stitching [6], that aim to minimize the global misalignment caused by parallax, medical image registration heavily relies on deformable registration and this motivates the current research focus [2][7]. For example, many medical image registration studies assume that the images have been affinely aligned after image preprocessing to remove global misalignment and thereby mainly focus on deformable registration with non-linear local deformations [8][9][10][11][12][13][14][15].

Deformable image registration aims to find a dense non-linear spatial correspondence (transformation) between a pair of fixed and moving images. Through the spatial transformation, the moving image can be warped to align with the fixed image. Traditional registration methods usually formulate deformable image registration as a time-consuming iterative optimization problem [16][17]. Recently, deep registration methods based on Deep Neural Networks (DNNs) have been widely used for their ability to perform fast end-to-end registration [3][7]. These methods learn a mapping from image pairs to spatial transformations based on training data, which have shown superior performance on both registration accuracy and speed when compared to traditional registration methods and hence are regarded as state-of-the-art methods [9][11][12][13][15][18].

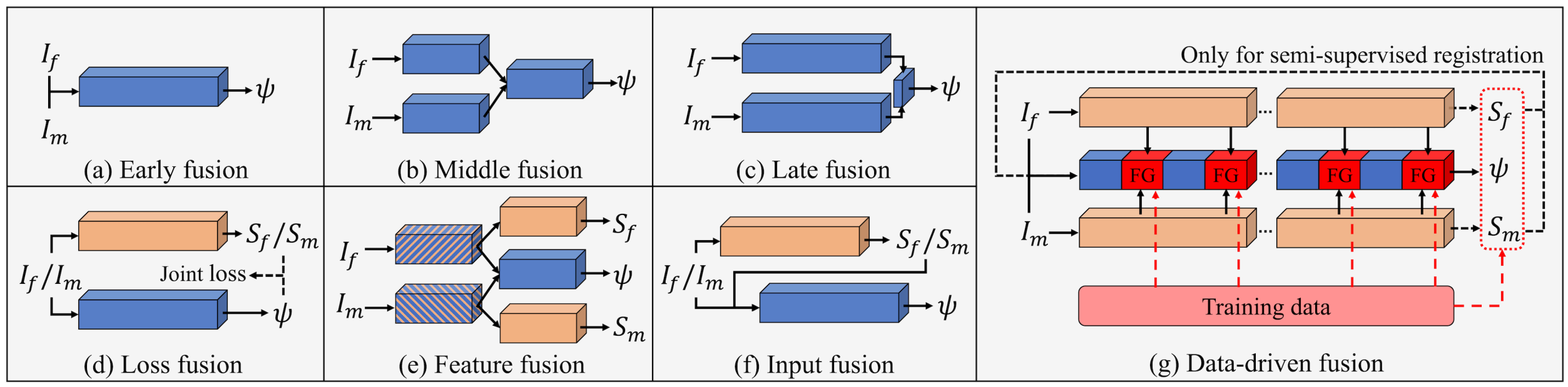

Fig. 1. Illustration of existing empirically-designed fusion strategies (a-f) and our data-driven fusion strategy (g) in unsupervised and semi-supervised settings. (a) Early fusion: $I_f$ and $I_m$ are concatenated as input. (b) Middle fusion: $I_f$ and $I_m$ enter separate encoders with intermediate features fused. (c) Late fusion: $I_f$ and $I_m$ enter separate networks with resultant features fused. (d) Loss fusion: $\psi$ and $S_f/S_m$ are mutually constrained by joint loss functions. (e) Feature fusion: multi-task networks are used with partial feature shared. (f) Input fusion: $S_f/S_m$ are fed as input for registration. (g) Data-driven fusion: fusion strategy is optimized during training. Legend: FG = Fusion Gate (FG) module; $I_f$ = fixed image; $I_m$ = moving image; $\psi$ = registration output; $S_f/S_m$ = segmentation output for $I_f/I_m$ (only for semi-supervised registration).

Early deep registration methods train DNNs in a fully supervised setting and require ground truth transformations between images as the labels [19][20]. However, ground truth transformations are unavailable. Therefore, imperfect transformation labels (estimated by traditional methods) or synthetic image pairs (with artificial transformations) are used as alternatives, which inevitably introduces inherited registration errors (from imperfect labels) or human experimental bias (from synthetic data) [13]. To remove the reliance on labels, recent deep registration methods have been developed to use image similarity metrics (e.g., mean square error) to train DNNs in a fully unsupervised setting [8][9][10][11][12][13][14][15][18]. In addition, weak anatomical labels (e.g., organ segmentation masks) have been used to complement the registration process and have been shown to improve registration performance [8]. Such weak labels enable DNNs to be trained in a semi-supervised setting where both unlabeled and weakly-labeled data are used [21][22][23][24][25][26][27][28].

To achieve end-to-end registration, deep registration methods need to explore the spatial information of each image and fuse this information to characterize spatial correspondence. Registration performance can be limited by poorly designed fusion strategies that hinder the characterization of spatial correspondence between images, while carefully designed fusion strategies can optimize the fusion of spatial information and greatly improve registration performance. This raises an essential question for image registration: *What is the optimal fusion strategy to characterize the spatial correspondence between images?* In addition, for semi-supervised registration with weak anatomical labels, anatomical segmentation can be jointly performed to improve image registration [21][22][24][25][26]. For this aim, semi-supervised deep registration methods also need to explore task-specific information for each task (registration and segmentation) and fuse this information to leverage the synergy between the two tasks. This raises another essential question: *What is the optimal fusion strategy to leverage the synergy between joint registration and segmentation?* For the first question, deep registration methods commonly adopt early fusion [8][9][10][11][14], middle fusion [12][29][30], or late fusion [31], as shown in Fig. 1(a-c). For the second question, semi-supervised deep registration methods usually adopt loss fusion [23][26][27], feature fusion [21][22][28], or input fusion [24][25], as shown in Fig. 1(d-f). Recent studies designed sophisticated fusion strategies and demonstrated that fusion strategies are a major influential factor to improve image registration [15][18][32][33][34][35][36][37]. Nevertheless, these sophisticated fusion strategies were empirically designed to fuse information by manually defined prior knowledge, which inevitably constrains the registration performance within the limits of empirical designs. The possible search space for fusion strategies is too large to be manually searched and the optimal fusion strategy could vary depending on data (e.g., medical images acquired from different scanners/organs or with different ranges of deformations), which inherently limits the development of empirically-designed fusion strategies.

In this study, we depart from the empirically-designed fusion strategies and develop a data-driven fusion strategy for deformable image registration, generalized to both unsupervised and semi-supervised registration. To achieve this, we propose (i) an Automatic Fusion network (AutoFuse) that provides flexibility to fuse information at many potential locations within the network, and (ii) a Fusion Gate (FG) module to control how to fuse information at each potential network location. As shown in Fig. 1(g), the FG modules enable our AutoFuse to optimize its fusion strategy based on training data during training, thus making a fundamental shift from existing empirically-designed fusion strategies to a data-driven fusion strategy. Moreover, our data-driven fusion strategy can be generalized to both Convolutional Neural Network (CNN) and transformer architectures and produce consistent improvements in the both CNN and transformer variants of AutoFuse. To the best of our knowledge, our AutoFuse is the first deep registration method that departs from empirically-designed fusion strategies and introduces a data-driven fusion strategy for both unsupervised and semi-supervised deformable image registration. Experiments on two well-benchmarked medical registration tasks (3D inter-patient brain image registration and 4D intra-patient cardiac image registration) with eight public datasets show that our AutoFuse outperforms the state-of-the-art unsupervised and semi-supervised registration methods.

## 2. Related Work

### 2.1. Unsupervised Medical Image Registration

Unsupervised registration methods are widely adopted as they do not require any data annotations. Traditional methods usually formulate deformable registration as an iterative optimization problem, which iteratively updates the spatial transformation for each image pair to maximize the similarity metrics between images [16][17]. To avoid the time-consuming iterative optimization in traditional methods, unsupervised deep registration methods have been proposed [8][9][10][11][12][13][14][15], in which DNNs were globally optimized to produce spatial transformations that maximize the similarity metrics of a set of training data and then were employed on unseen testing data without the need for further optimization.

Most existing unsupervised deep registration methods adopted the basic early, middle, or late fusion strategies [8][9][10][11][12][14][31]. Recently, some sophisticated fusion strategies have also been used to improve unsupervised registration [15][18][32][33][34][36]. Chen et al.[32] proposed to use cross-stitch units [38] to extract and fuse features for affine registration. Also for affine registration, Chen et al.[33] proposed a dual-channel squeeze-fusion-excitation co-attention module to fuse information. For deformable registration, Zhang et al.[18] proposed a Dual Transformer Network (DTN) that adopted both early and middle fusion strategies. Shi et al.[15] used a transformer network (named XMorpher) that includes dual parallel feature extraction networks with information fused by cross-attention. More recently, Ma et al.[36] proposed a Separate Encoding Network (SEN) that includes three parallel encoders to extract features (from separate images and concatenated image pairs) and fuse these features at multiple scales. Chen et al.[34] proposed a dual-stream transformer-based network (named TransMatch) that extracts features through symmetrical dual encoders with self-attention and then adopts cross-attention to realize feature matching and fusion. However, as we have mentioned, these sophisticated fusion strategies are empirically-designed and inherently limited.

### 2.2. Semi-supervised Medical Image Registration

In addition to fully unsupervised registration, anatomical information has been introduced to improve registration. As anatomical segmentation labels can be used to evaluate whether images are aligned well in anatomy, segmentation labels hence can be regarded as weak labels for registration. Early weakly-supervised deep registration methods leveraged the weak labels to directly supervise the network's training and achieved better performance than their unsupervised counterparts [8][29][39]. However, these methods required segmentation labels to be available for all training data and thus cannot leverage the easily accessible unlabeled data as an enhancement for training. To ease the reliance on segmentation labels, semi-supervised deep registration methods were proposed, where only a small number of weak labels together with unlabeled data are required for training [21][22][23][24][25][26][27][28].

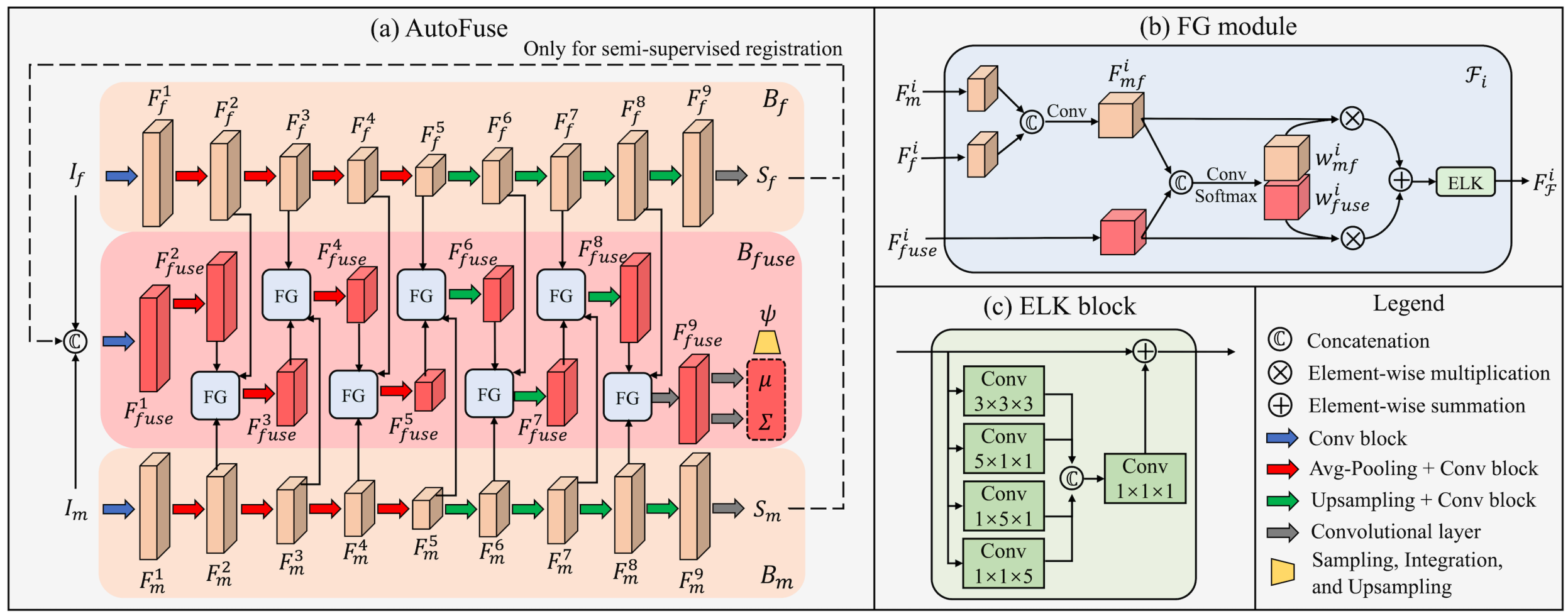

Fig. 2. Overview of the proposed Automatic Fusion network (AutoFuse), including the architecture of (a) AutoFuse, (b) Fusion Gate (FG) modules, and (c) Efficient Large Kernel (ELK) blocks. Note that the skip connections of each branch are omitted in this figure for the sake of the clarity. The branches $B_m$ and $B_f$ share the same weights.

Semi-supervised registration methods usually perform joint registration and segmentation to leverage their synergy. For this aim, existing semi-supervised methods commonly adopted loss, feature, or input fusion strategies. Loss fusion is crucial and almost all semi-supervised methods adopted this strategy to some extent [21][22][23][24][25][26][27][28]. With loss fusion, the segmentation outputs can serve as weak anatomical labels to constrain the registration outputs, while the registration outputs can augment segmentation labels to constrain the segmentation outputs. The loss fusion strategy can be adopted alone [23][26][27] or jointly with the feature fusion strategy [21][22][28] or input fusion strategy [24][25]. Recently, there have been sophisticated fusion strategies proposed for semi-supervised registration [35][37]. Khor et al.[35] proposed a deformable registration network (named AC-DMiR) that realizes anatomically-constrained attention-guided feature fusion to maximize the information flow between the registration and segmentation tasks. Ma et al.[37] proposed a global-local transformation network (GL-Net) that implements both input and loss fusion through a region similarity constraint. However, these fusion strategies for semi-supervised registration are also empirically-designed and inherently limited.

## 3. Method

Image registration aims to find a spatial transformation $\psi$ that warps a moving image $I_m$ to a fixed image $I_f$, so that the warped image $I_{m\circ\psi} = I_m \circ \psi$ is spatially aligned with the fixed image $I_f$. In this study, we assume that the $I_m$ and $I_f$ are two single-channel, grayscale volumes defined in a 3D spatial domain $\Omega \subset \mathbb{R}^3$, which is consistent with common medical image registration studies [8][9][11][12][13][14][15][18][29][34][36]. The $\psi$ is parameterized as a diffeomorphic deformation field to ensure the invertibility and topology-preservation of spatial transformations [10][14][15][16]. We parametrized the deformable registration problem as a function $\mathcal{R}_\theta(I_f, I_m) = \psi$ using our AutoFuse (detailed in Section 3.1). The $\theta$ is a set of learnable parameters that can be learned in the fully unsupervised setting (detailed in Section 3.2) or in the semi-supervised setting (detailed in Section 3.3).

### 3.1. Automatic Fusion Networks (AutoFuse)

Fig. 2 shows the architecture of our AutoFuse, which consists of three branches with Unet-style encoder-decoder structure. Two branches, denoted by $B_m$ and $B_f$, extract features from $I_m$ and $I_f$ separately, while the other branch, denoted by $B_{fuse}$, first extracts features from concatenated $I_m$ and $I_f$ and then fuse the features from $B_m$ and $B_f$ via FG modules (detailed in Section 3.1.1).

Each branch consists of successive Conv blocks and each Conv block is composed of two 3×3×3 convolutional layers followed by LeakyReLU activation (with parameter 0.2) and instance normalization. The encoder uses average pooling layers to reduce the resolution of feature maps, while the decoder uses upsampling layers to increase the resolution of feature maps. FG modules are embedded after the Conv blocks of $B_{fuse}$ for feature fusion. Formally, let $F_m^i$, $F_f^i$, and $F_{fuse}^i$ be the features from the $i^{th}$ Conv block of $B_m$, $B_f$, and $B_{fuse}$. Let $\mathcal{F}_i$ be the FG module after the $i^{th}$ Conv block and $F_{\mathcal{F}}^i$ be the fused features from the $\mathcal{F}_i$. We can derive $F_{\mathcal{F}}^i = \mathcal{F}_i(F_m^i, F_f^i, F_{fuse}^i)$ and the $F_{\mathcal{F}}^i$ will be fed into the next Conv block of $B_{fuse}$ for later fusion. Skip connections are used between the encoder and decoder of each branch. For $B_{fuse}$, the outputs of FG modules are propagated to the decoder through skip connections. The $B_M$ and $B_F$ share weights and their kernel numbers are half of the $B_{fuse}$. The convolutional kernel numbers used in our experiments are presented in Appendix A.

To obtain the diffeomorphic deformation field $\psi$, the output of $B_{fuse}$ is fed into two parallel convolutional layers to produce a mean map $\mu$ and a variance map $\Sigma$. Then, a stationary velocity field $v$ is sampled from the $\mu$ and $\Sigma$, and is converted into the $\psi$ through seven steps of scaling and squaring integration [10]. Following [10], the feature maps of $B_{fuse}$ are unsampled three times in the decoder, and the sampling and integration operations are performed at half of the original scale, which ensures the network can fit into the GPU memory. Accordingly, the FG module is also not used at the original scale.

When our AutoFuse is employed for semi-supervised registration, the outputs of $B_m$ and $B_f$ are fed into a softmax-activated convolutional layer to produce segmentation masks $S_m$ and $S_f$. The $S_m$ and $S_f$ are propagated back and concatenated with the $I_m$ and $I_f$ as the input of $B_{fuse}$.

### 3.1.1 Fusion Gate (FG) Module

To selectively fuse the features from different branches, we propose the FG module based on attention mechanisms. As shown in Fig. 2(b), for the FG module $\mathcal{F}_i$, the $F_m^i$ and $F_f^i$ are first fused as $F_{mf}^i$ using a 3×3×3 convolutional layer. Then, the $F_{mf}^i$ and $F_{fuse}^i$ are concatenated to learn two adaptive weight maps $w_{mf}^i$ and $w_{fuse}^i$ by two 1×1×1 convolutional layers and Softmax function. With the $w_{mf}^i$ and $w_{fuse}^i$, the weighted summation of $F_{mf}^i$ and $F_{fuse}^i$ is fed into an Efficient Large Kernel (ELK) block (detailed in Section 3.1.2) for feature refinement. Finally, we can derive $F_{\mathcal{F}}^i = ELK(w_{mf}^i F_{mf}^i + w_{fuse}^i F_{fuse}^i)$ as the output of $\mathcal{F}_i$.

Through the proposed FG module, our AutoFuse can automatically optimize its fusion strategy during training. All FG modules serve as gates to control how to fuse information in the AutoFuse. In the unsupervised setting, early, middle, and late fusion can be potentially employed as needed. For example, if all the learned $w_{mf}^i$ are zero maps, the $F_{\mathcal{F}}^i$ will be fully determined by $F_{fuse}^i$ and the AutoFuse will employ early fusion; if the learned $w_{mf}^i$ are one maps in the encoder and zero maps in the decoder, the AutoFuse will employ middle fusion. However, as the learned $w_{mf}^i$ and $w_{fuse}^i$ are not explicitly encouraged to be zero or one maps, the AutoFuse usually will employ all early, middle, and late fusion but impose different weights. Similarly in the semi-supervised setting, loss, feature, and input fusion also can be potentially employed as needed. We expect our AutoFuse can search over a large possible space of fusion strategies and finally find an optimal strategy based on training data.

### 3.1.2 Efficient Large Kernel (EKL) Block

The ELK block is a memory-efficient variant of the Large Kernel (LK) block [11]. Jia et al.[11] employed large kernel (5×5×5) convolution in LK blocks to increase the effective receptive field of a vanilla U-Net [40] and showed that the U-Net with LK blocks (LKU-Net) can outperform the recent state-of-the-art transformer-based registration method, TransMorph [9]. We followed Jia et al.[11] but modified the original LK block into the memory-efficient ELK block. To reduce memory consumption, the original large kernel convolutional layer is replaced by three parallel convolutional layers with large kernels only in one direction. As shown in Fig. 2(c), there are four parallel convolutional layers with a kernel size of 3×3×3, 5×1×1, 1×5×1, and 1×1×5 in each ELK block. The outputs of these four parallel layers are concatenated and fed into a 1×1×1 convolutional layer for integration. An identity shortcut is used between the input and output of each ELK block.

### 3.1.3 Transformer-based Variant

Transformers have been widely adopted in many medical image applications for their capabilities to capture long-range dependency [41]. Recently, transformers have also been used for image registration and shown superior performance than their CNN counterparts [9]. To explore the generalizability of the proposed data-driven fusion strategy in transformer-based networks, we propose a transformer-based variant of AutoFuse, named AutoFuse-Trans, by replacing its Conv blocks with Swin transformer blocks [42]. The first Conv block of each branch is replaced by a patch embedding layer with patch size of 2×2×2 [42], where the image input is converted into sequence embeddings at half of the original image scale and enter the following swin transformer blocks. For feature downsampling and upsampling, the average pooling layers and upsampling layers are replaced by patch merging layers and patch expanding layers [43]. The detailed architectural settings, including window size, embedding dimensions, and attention head numbers, are presented in Appendix A.

## 3.2. Unsupervised Learning

In the unsupervised setting, the learnable parameters $\theta$ of our AutoFuse are optimized using an unsupervised loss $\mathcal{L}_{uns}$ that does not require labels. The $\mathcal{L}_{uns}$ consists of two terms $\mathcal{L}_{sim}$ and $\mathcal{L}_{reg}$, where the $\mathcal{L}_{sim}$ is an image similarity term that penalizes the differences between the warped image $I_{m\circ\psi}$ and the fixed image $I_f$, while the $\mathcal{L}_{reg}$ is a regularization term that encourages smooth and invertible diffeomorphic transformations $\psi$.

For the $\mathcal{L}_{sim}$, we adopt negative local normalized cross-correlation (NCC), a similarity metric that has been widely used in deformable image registration methods [8][9][10][11][12][13][14][18]. For the $\mathcal{L}_{reg}$, we first adopt the Kullback-Leibler divergence (KL) between the true and approximate posteriors based on the predicted mean map $\mu$ and variance map $\Sigma$ [10], in which a smoothing precision parameter $\lambda$ is used to balance the registration accuracy and transformation smoothness. In addition, as the $\psi$ is not invertible at the voxel $p$ where the Jacobian determinant is negative (i.e., $|J\psi(p)| \leq 0$) [60], we also adopt a Jacobian Determinant-based loss (JD) [5] to explicitly penalize the negative Jacobian determinants of $\psi$. Consequently, the $L_{uns}$ is finally defined as:

$$\mathcal{L}_{uns} = -NCC\left(I_f, I_{m\circ\psi}\right) + \sigma KL_{\lambda}(\mu, \Sigma) + \mu JD(\psi), \tag{1}$$

where the $\sigma$, $\lambda$ and $\mu$ are three regularization parameters. In the experiments, we only adjusted $\lambda$ and $\mu$ but fixed $\sigma$ as 0.01 to make the value of KL term close to the NCC term.

## 3.3. Semi-supervised Learning

In the semi-supervised setting, the learnable parameters $\theta$ of our AutoFuse are optimized using a semi-supervised loss $\mathcal{L}_{semi}$ that requires segmentation labels available for partial training data. The $L_{semi}$ is defined as:

$$\mathcal{L}_{semi} = \mathcal{L}_{uns} + \alpha\mathcal{L}_{seg} + \beta\mathcal{L}_{fuse}, \tag{2}$$

where the $\mathcal{L}_{seg}$ is a segmentation term that constrains the predicted segmentation masks $S_m$ and $S_f$, while the $\mathcal{L}_{fuse}$ is a fusion term that imposes mutual constraints between registration and segmentation (i.e., loss fusion). The $\alpha$ and $\beta$ are two balancing parameters that were set to be 1 as default in our experiments.

For the $\mathcal{L}_{seg}$, we calculate the sum of Dice [44] and Focal [45] losses (denoted by $FocalDice$). Let $L_m$ and $L_f$ be the segmentation labels of $I_m$ and $I_f$. The $\mathcal{L}_{seg}$ is defined as:

$$\mathcal{L}_{seg} = FocalDice(L_m, S_m) + FocalDice\left(L_f, S_f\right). \tag{3}$$

For the $\mathcal{L}_{fuse}$, we warp the $S_m$ and $L_m$ to be $S_{m\circ\psi} = S_m \circ \psi$ and $L_{m\circ\psi} = L_m \circ \psi$, and calculate the Dice and Focal losses between the warped and fixed segmentation masks. The $\mathcal{L}_{fuse}$ is defined as:

$$\mathcal{L}_{fuse} = FocalDice\left(L_f, S_{m\circ\psi}\right) + FocalDice\left(S_f, L_{m\circ\psi}\right) + FocalDice\left(L_f, L_{m\circ\psi}\right). \quad (4)$$

The $\mathcal{L}_{semi}$ can adapt to semi-supervised image pairs that consist of a labeled image and an unlabeled image. For example, if the $L_m$ is unavailable and the $L_m$-related terms in Eq.(3) and Eq.(4) are invalid (not calculated), the $L_f$ still can serve as a coarse segmentation label to constrain the $S_m$, while the $S_m$ can serve as an anatomical label (with the $L_f$) to constrain the $\psi$. With this design, both labeled and unlabeled data can be effectively used for training.

## 4. Experimental Setup

### 4.1. Datasets and Preprocessing

We evaluated our AutoFuse with two well-benchmarked medical registration tasks (3D inter-patient brain image registration and 4D intra-patient cardiac image registration), which involved a total of eight public medical image datasets:

For inter-patient brain image registration, we adopted seven public 3D brain Magnetic Resonance Imaging (MRI) image datasets that have been widely used for brain image registration evaluation [3][7][46]. We collected 414 brain MRI images with segmentation labels from OASIS [47] and randomly split them into 314, 20, and 80 images for training, validation, and testing. We also collected 2,656 unlabeled brain MRI images from four public datasets, ADNI [48], ABIDE [49], ADHD [50], and IXI [51], and used them for training. This results in a large semi-supervised training set consisting of 2,656 unlabeled and 314 labeled images. In addition, we used two public brain MRI datasets with segmentation labels, Mindboggle [52] and Buckner [53], for independent testing. The Mindboggle and Buckner datasets contain 100 and 40 images, which were merely used for testing and fully independent from training and validation. A total of 35, 62, and 110 anatomical structures were segmented as labels in the OASIS, Mindboggle, and Buckner datasets. In the unsupervised setting, the segmentation labels were fully independent from the training process and were merely used for evaluation. In the semi-supervised setting, we preprocessed the OASIS segmentation labels to reduce the label channels for more efficient training, where the symmetric anatomical structures in the left and right brain hemispheres were merged following [27] and this resulted in 19 remaining anatomical structures. Examples of segmentation labels are provided in Fig. 3, which delineate the labeled anatomical structures with different colors. Note that only the preprocessed OASIS labels are available for a small proportion (~10%) of training set, while the registration performance is evaluated with the original segmentation labels on three different testing sets (OASIS, Mindboggle, and Buckner). We followed the existing literatures [8][9][10][11][15][18] and performed inter-patient registration for evaluation, where we randomly picked 100 image pairs from each of the OASIS, Mindboggle, and Buckner testing sets, resulting in a total of 300 testing image pairs. We performed standard brain MRI preprocessing steps, including brain extraction, intensity normalization, and affine registration, with FreeSurfer [53] and FLIRT [54]. All images were affine-transformed and resampled to align with the MNI-152 brain template [55] with 1mm$^3$ isotropic voxels, and then were cropped into 144×192×160.

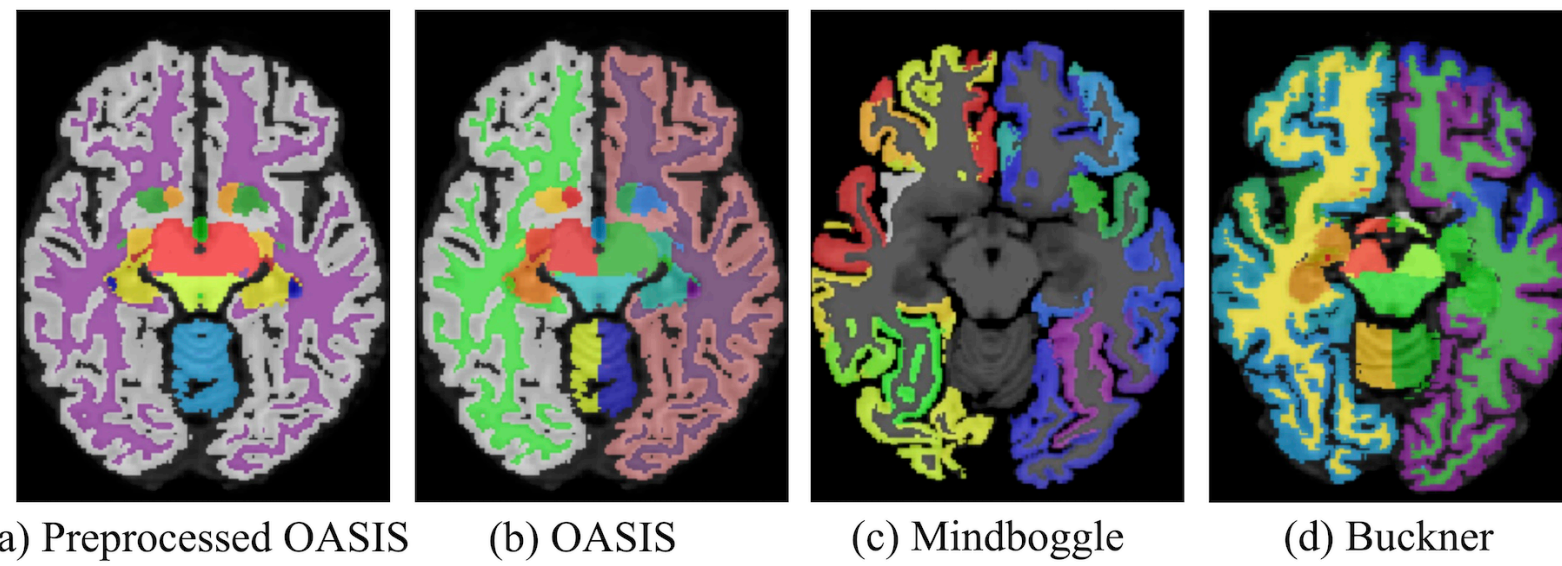

Fig. 3. Examples of the segmentation labels used for semi-supervised learning (a) and evaluation on the OASIS (b), Mindboggle (c), and Buckner (d) datasets. The labeled anatomical structures are delineated with different colors. 2D cross-sectional slices are visualized for illustration.

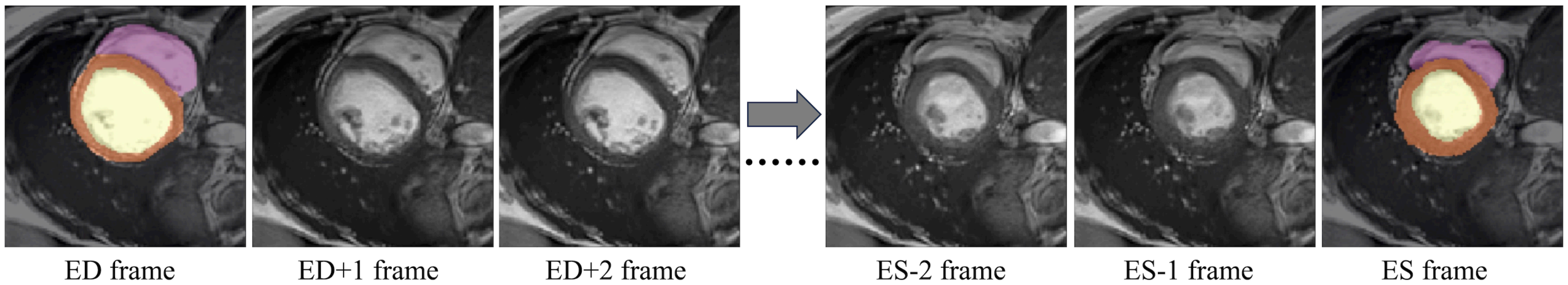

Fig. 4. Examples of the cine-MRI images in the ACDC dataset. From left to right are the ED frame, the intermediate frames from ED to ES, and the ES frame. The three labeled anatomical regions are colored in the ED and ES frames. 2D cross-sectional slices are visualized for illustration.

For intra-patient cardiac image registration, we adopted the public ACDC dataset [56] that contains 4D cardiac cine-MRI images of 150 patients. Each 4D cine-MRI image contains tens of 3D MRI frames acquired from different time-points including the End-Diastole (ED) and End-Systole (ES) frames. The ED was defined as the first frame when the mitral valve was closed and the ES was defined as the first frame when the aortic valve was closed. ED and ES delineate the two ends of cardiac cycle and show the largest deformation in a cardiac cycle [57]. The ACDC dataset provides 100 cine-MRI images in the training set and 50 cine-MRI images in the testing set, where we further randomly divided the training set into 90 and 10 cine-MRI images for training and validation and used the provided testing set for testing. Three anatomical regions, including left ventricular cavity, right ventricular cavity, and myocardium, were segmented as labels in the ED and ES frames of each cine-MRI image, resulting in a semi-supervised dataset consisting of labeled ED/ES frames and unlabeled non-ED/ES frames. The examples of cardiac cine-MRI are provided in Fig. 4, where the three labeled anatomical regions are colored in the ED and ES frames. The segmentation labels were used for semi-supervised learning and evaluation, which were fully independent from the training process in the unsupervised setting. Following Krebs et al.[58] and Kim et al.[59], we aim to register the ED and ES frames of the same patient. The intra-patient ED and ES frames were registered with each other (ED-to-ES and ES-to-ED), where 100 testing image pairs were derived from the testing set. All cine-MRI frames were resampled with a voxel spacing of 1.5×1.5×3.15mm$^3$ and cropped to 128×128×32 voxels around the center. The voxel intensity was normalized to range [0, 1] through max-min normalization.

### 4.2. Implementation Details

We implemented the AutoFuse using PyTorch on a NVIDIA V100 GPU with 32 GB memory. We used an ADAM optimizer with a learning rate of 0.0001 and a batch size of 1. We set $\lambda = 50$ and $\mu = 10^{-5}$ to ensure that the percentage of voxels with negative Jacobian determinants is less than 0.05% (refer to the regularization analysis in Section 5.1.3). Our code is publicly available at https://github.com/MungoMeng/Registration-AutoFuse.

For inter-patient brain image registration, the AutoFuse was trained for a total of 100,000 iterations with inter-patient image pairs. In the unsupervised setting, the training image pairs were randomly picked from the training set. In the semi-supervised setting, the AutoFuse was first trained for 50,000 iterations using labeled images as a warm start, and then was trained for another 50,000 iterations with semi-supervised image pairs that consist of a labeled image and an unlabeled image, thus enhancing the model's generalizability by incorporating unlabeled images into training. Validation was performed after every 1,000 training iterations and the model achieving the highest validation result was preserved for final testing.

For intra-patient cardiac image registration, the AutoFuse was trained for a total of 50,000 iterations with intra-patient image pairs. In the unsupervised setting, the AutoFuse was first trained for 40,000 iterations with image pairs randomly picked from the training set (including both ED/ES and other frames), and then was trained for another 10,000 iterations with ED/ES image pairs consisting of ED/ES frames only, thus optimizing the model to register ED/ES frames. In the semi-supervised setting, the AutoFuse was first trained for 40,000 iterations with semi-supervised image pairs that consist of a labeled ED/ES frame and an unlabeled non-ED/ES frame, and then was also trained for another 10,000 iterations with ED/ES image pairs to optimize ED/ES registration.

### 4.3. Comparison Methods

Our AutoFuse was extensively compared to the state-of-the-art registration methods. In the unsupervised setting, two traditional methods and eight deep registration methods were compared. The two traditional methods are SyN [16] and NiftyReg [17], and we ran them using cross-correlation as the similarity measure. The eight deep registration methods are VoxelMorph (VM) [8], Diffeomorphic VoxelMorph (DifVM) [10], LKU-Net [11], TransMorph [9], DTN [18], XMorpher [15], SEN [36], and TransMatch [34]. The VM and DifVM are two commonly benchmarked methods for medical image registration [9][11][13][15][18]. The LKU-Net and TransMorph are two state-of-the-art methods that employ LK blocks and transformers to improve VM. The DTN, XMorpher, SEN, and TransMatch are four recent methods that employ sophisticated fusion strategies, which have been discussed in Section 2.1.

In the semi-supervised setting, six deep registration methods were compared, including weakly-supervised VM (WS-VM) [8], RSegNet [26], JPR-Net [28], PC-Reg [24], AC-DMiR [35], and GL-Net [37]. The WS-VM is the weakly-supervised variant of VM, which was trained only with labeled data. The RSegNet is a semi-supervised method that employs loss fusion alone, while the JPR-Net and PC-Reg further incorporated feature fusion and input fusion, respectively. The AC-DMiR and GL-Net are two recent semi-supervised methods that employs sophisticated fusion strategies as discussed in Section 2.2.

We followed the corresponding references to implement the deep registration methods with two modifications: (i) We adopted NCC as the similarity loss for all deep registration methods for a fair comparison, and (ii) The JPR-Net was modified to adapt to 3D images, where spherical operations were replaced by 3D operations while network topology was unchanged.

### 4.4. Experimental Settings

Our AutoFuse was compared with the existing unsupervised and semi-supervised registration methods for both brain and cardiac image registration. We adopted standard evaluation metrics for medical image registration [5][8][9][10][11][13][15][18][]: The registration accuracy was evaluated using the Dice similarity coefficients (DSC) between the segmentation labels of the warped and fixed images, while the smoothness and invertibility of spatial transformations were evaluated using the percentage of Negative Jacobian Determinants (NJD). Generally, a higher DSC and a lower NJD indicate a better registration performance. A two-sided $P$ value less than 0.05 is considered to indicate a statistically significant difference between two average DSC values.

In the unsupervised setting, we performed an ablation study where our AutoFuse was compared to baseline methods that employ the basic early, middle, and late fusion strategies (denoted by EarlyFuse, MidFuse, and LateFuse). The EarlyFuse, MidFuse, and LateFuse have the same encoder-decoder architecture as AutoFuse but use different fusion strategies. We also built a baseline method that excludes FG modules and fuses image features at multiple scales (denoted by MultiFuse). The MultiFuse directly sums $F_{mf}^{i}$ and $F_{fuse}^{i}$ without using the FG modules. For a fair comparison, we purposely increase the kernel numbers of each baseline method, so that all the baseline methods have similar or higher parameter numbers than the AutoFuse. The detailed architectural settings of these baseline methods are provided in Appendix C. In addition, we performed a regularization analysis on the parameters $\lambda$ and $\mu$ to explore the trade-off between registration accuracy (DSC) and transformation invertibility (NJD). For comparison, we also implemented an AutoFuse variant parameterizing the $\psi$ as a displacement field (denoted by AutoFuse-disp), which excludes the diffeomorphic constraints and the KL loss term in Eq. (1).

In the semi-supervised setting, we also performed an ablation study where feature fusion, input fusion, or FG modules was removed in baseline methods. To achieve this, we removed the relevant connections or layers but did not alter the overall topology of AutoFuse. The detailed architectural settings of these baseline methods are provided in Appendix C. In addition, we performed a semi-supervised learning analysis to explore the impacts of data annotations on registration performance, in which varying numbers of labeled images were used for semi-supervised learning. For the auxiliary segmentation task, the performance of our AutoFuse is reported quantitatively and qualitatively in Appendix B.

Furthermore, we performed a qualitative comparison between our AutoFuse and the existing registration methods in both unsupervised and semi-supervised settings. We also provided a statistical interpretation on the proposed data-driven fusion strategy by calculating the mean values of the adaptive weight maps in FG modules.

## 5. Results

### 5.1. Unsupervised Registration Evaluation

#### 5.1.1 Comparison with Existing Methods

Table 1 shows the quantitative comparison between the AutoFuse and existing registration methods for inter-patient brain image registration in the unsupervised setting. We report the DSC and NJD results on three testing sets that involve different anatomical structures for evaluation (OASIS, Mindboggle, and Buckner). Our AutoFuse achieved significantly higher DSC results ($P$<0.05) than all the comparison methods across the three testing sets. Our AutoFuse-Trans further improved the DSC results over the original AutoFuse, which outperformed the comparison methods by a larger margin. As a diffeomorphic registration method, our AutoFuse also achieved the best NJD results. The compared diffeomorphic methods (DifVM and DTN) also achieved similar NJD results to our AutoFuse, but their DSC results were significantly worse ($P$<0.05). Compared with diffeomorphic methods, non-diffeomorphic methods (VM, TransMorph, LKU-Net, SEN, XMorpher, and TransMatch) achieved competitive performance in DSC, but their NJD results were obviously worse (~2.0% vs ~0.05%). In addition, the runtime results show that our AutoFuse is much faster than the traditional registration methods (SyN and NiftyReg) while being similar to the existing deep registration methods.

Table 2 shows the quantitative comparison between the AutoFuse and existing registration methods for intra-patient cardiac image registration in the unsupervised setting. For cardiac image registration, diffeomorphic methods (DifVM and DTN) showed advantages in both DSC and NJD. For example, the DifVM achieved better DSC and NJD results than its non-diffeomorphic counterpart (VM). Our AutoFuse and AutoFuse-Trans, as diffeomorphic registration methods, also showed advantages in NJD and achieved significantly higher DSC results ($P$<0.05) than all the comparison methods.

Table 1. Quantitative comparison for unsupervised brain image registration.

| Method | OASIS (%) | | Mindboggle (%) | | Buckner (%) | | Runtime (second) | |
|---|---|---|---|---|---|---|---|---|
| | DSC ↑ | NJD ↓ | DSC ↑ | NJD ↓ | DSC ↑ | NJD ↓ | CPU ↓ | GPU ↓ |
| Before registration | 61.0* | / | 34.7* | / | 40.6* | / | / | / |
| SyN | 76.2* | 0.204 | 52.8* | 0.216 | 56.4* | 0.147 | 3427 | / |
| NiftyReg | 78.7* | 0.246 | 56.7* | 0.264 | 61.0* | 0.188 | 159 | / |
| VM | 77.3* | 1.997 | 55.2* | 2.532 | 58.9* | 2.220 | **2.84** | **0.32** |
| DifVM | 78.4* | 0.041 | 52.8* | 0.043 | 57.4* | 0.039 | 2.92 | 0.34 |
| LKU-Net | 79.3* | 1.792 | 57.4* | 2.217 | 61.2* | 1.992 | 3.34 | 0.42 |
| TransMorph | 79.2* | 1.908 | 57.1* | 2.400 | 60.8* | 2.183 | 3.68 | 0.45 |
| DTN | 78.9* | 0.060 | 56.1* | 0.060 | 60.1* | 0.056 | 3.31 | 0.41 |
| XMorpher | 79.0* | 1.688 | 57.0* | 2.228 | 61.0* | 1.800 | 4.18 | 0.47 |
| SEN | 78.2* | 1.976 | 56.2* | 2.370 | 60.4* | 2.115 | 3.17 | 0.36 |
| TransMatch | 79.8* | 1.411 | 57.6* | 1.520 | 62.0* | 1.306 | 3.06 | 0.35 |
| AutoFuse (ours) | 80.8 | 0.025 | 59.0 | 0.031 | 63.5 | 0.024 | 3.26 | 0.38 |
| AutoFuse-Trans (ours) | **81.3** | **0.024** | **59.8** | **0.029** | **64.1** | **0.022** | 5.87 | 0.65 |

**Bold**: the best result in each column. *: $P$<0.05, in comparison to AutoFuse. ↑: the higher is better. ↓: the lower is better.

Table 2. Quantitative comparison for unsupervised cardiac image registration.

| Method | ACDC (%) | | Runtime (second) | |
|---|---|---|---|---|
| | DSC ↑ | NJD ↓ | CPU ↓ | GPU ↓ |
| Before registration | 59.0* | / | / | / |
| SyN | 74.7* | 0.154 | 401 | / |
| VM | 75.4* | 0.440 | **0.36** | **0.04** |
| DifVM | 77.3* | 0.051 | 0.49 | 0.06 |
| LKU-Net | 77.0* | 0.427 | 0.80 | 0.08 |
| TransMorph | 76.9* | 0.497 | 0.89 | 0.09 |
| DTN | 77.8* | 0.088 | 0.77 | 0.08 |
| XMorpher | 76.5* | 0.353 | 1.25 | 0.11 |
| SEN | 75.9* | 0.452 | 0.43 | 0.06 |
| TransMatch | 77.0* | 0.256 | 0.40 | 0.06 |
| AutoFuse (ours) | 79.6 | **0.036** | 0.72 | 0.07 |
| AutoFuse-Trans (ours) | **80.2** | 0.045 | 1.66 | 0.15 |

**Bold**: the best result in each column. *: *P*<0.05, in comparison to AutoFuse.
↑: the higher is better. ↓: the lower is better. The results of NiftyReg are missing as it failed to converge when applied to cardiac image registration.

### 5.1.2 Ablation Study

Table 3 shows the DSC results of the ablation study for inter-patient brain image registration in the unsupervised setting. The NJD results are omitted as all methods adopted the same regularization settings and achieved similar NJD results. The parameter numbers are also reported in Table 3, where the four baseline methods have similar or higher parameter numbers than our AutoFuse. Among the baseline methods, the MultiFuse achieved the highest DSC results, followed by MidFuse, EarlyFuse, and LateFuse. Our AutoFuse outperformed the MultiFuse and achieved significantly higher DSC results (*P*<0.05) than all the baseline methods. Note that ELK blocks were not used by AutoFuse in this ablation study for a fair comparison.

Table 3. DSC results of the ablation study for unsupervised brain image registration.

| Method | Parameter Num | OASIS (%) | Mindboggle (%) | Buckner (%) |
|---|---|---|---|---|
| EarlyFuse | 8.06M | 78.2* | 56.0* | 60.3* |
| MidFuse | 8.17M | 78.3* | 56.2* | 60.5* |
| LateFuse | 8.17M | 77.6* | 55.3* | 59.4* |
| MultiFuse | 9.71M | 78.6* | 56.5* | 61.1* |
| AutoFuse‡ (ours) | 8.75M | **80.0** | **58.5** | **62.9** |

**Bold**: the highest DSC in each column. *: *P*<0.05, in comparison with AutoFuse. ‡: without ELK blocks.

### 5.1.3 Regularization Analysis

Table 4 shows the validation results of the AutoFuse using different regularization settings for inter-patient brain image registration in the unsupervised setting. The AutoFuse-disp with $\mu = 0$ did not impose any explicit constraints on the negative Jacobian determinants and obtained the worst NJD. Using the JD loss (set $\mu = 10^{-5}$) enabled the AutoFuse-disp to achieve a better NJD with a slightly degraded DSC. By incorporating diffeomorphic constraints, our AutoFuse showed the potential to outperform the baseline AutoFuse-disp on both DSC and NJD, but its DSC results were dramatically degraded as the $\lambda$ increased: The AutoFuse with $\lambda = 50$ and $\mu = 0$ achieved the best DSC result with a NJD > 1.0%, while the AutoFuse with $\lambda = 400$ and $\mu = 0$ achieved the worst DSC result with a NJD < 0.05%. Our AutoFuse with $\lambda = 50$ and $\mu = 10^{-5}$ achieved the overall best validation results, which obtained the second-highest DSC with a NJD < 0.05%. Compared to increasing the $\lambda$, adding the JD loss (set $\mu = 10^{-5}$) enabled our AutoFuse to achieve better NJD with a smaller decrease in DSC.

Table 4. Validation results of the AutoFuse using different regularization settings on the OASIS validation set.

| Method | | DSC (%) ↑ | NJD (%) ↓ |
|---|---|---|---|
| AutoFuse-disp | $\mu = 0$ | 80.8 | 1.741 |
| | $\mu = 10^{-5}$ | 80.5 | 0.059 |
| AutoFuse (ours) | $\lambda = 50, \mu = 0$ | **81.4** | 1.074 |
| | $\lambda = 200, \mu = 0$ | 80.9 | 0.223 |
| | $\lambda = 400, \mu = 0$ | 80.2 | **0.036** |
| | $\lambda = 50, \mu = 10^{-5}$ | 81.0 | 0.038 |

**Bold**: the best result in each column. ↑: the higher is better. ↓: the lower is better.

## 5.2. Semi-supervised Registration Evaluation

### 5.2.1 Comparison with Existing Methods

Table 5 shows the quantitative comparison between the AutoFuse and existing registration methods for inter-patient brain image registration in the semi-supervised setting. Consistent with the unsupervised results in Table 1, our AutoFuse also achieved the best DSC and NJD results for semi-supervised registration across three testing sets (OASIS, Mindboggle, and Buckner). Compared to the unsupervised setting, our AutoFuse gained 3.7%, 1.2%, and 1.5% DSC improvements ($P$<0.05) in the semi-supervised setting, which enabled our AutoFuse to achieve significantly higher DSC results ($P$<0.05) than the compared semi-supervised registration methods. In addition, our AutoFuse-Trans also improved the DSC results over the original AutoFuse in the semi-supervised setting and thus outperformed the comparison methods by a larger margin. The runtime results show that our AutoFuse required similar runtime to the state-of-the-art semi-supervised deep registration methods.

Table 6 shows the quantitative comparison between the AutoFuse and existing registration methods for intra-patient cardiac image registration in the semi-supervised setting. Also consistent with the unsupervised results in Table 2, our AutoFuse and AutoFuse-Trans achieved the best DSC and NJD results for intra-patient cardiac image registration in the semi-supervised setting, which showed significantly higher DSC results ($P$<0.05) than all the comparison methods. Compared to the unsupervised setting, our AutoFuse gained 6.1% DSC improvement ($P$<0.05) for cardiac image registration in the semi-supervised setting.

Table 5. Quantitative comparison for semi-supervised brain image registration.

| Method | OASIS (%) | | Mindboggle (%) | | Buckner (%) | | Runtime (second) | |
|---|---|---|---|---|---|---|---|---|
| | DSC ↑ | NJD ↓ | DSC ↑ | NJD ↓ | DSC ↑ | NJD ↓ | CPU ↓ | GPU ↓ |
| Before registration | 61.0* | / | 34.7* | / | 40.6* | / | / | / |
| WS-VM | 79.4* | 1.844 | 56.0* | 2.337 | 59.4* | 2.038 | **2.85** | **0.32** |
| RSegNet | 81.6* | 0.035 | 56.3* | 0.038 | 60.2* | 0.031 | 3.02 | 0.34 |
| JPR-Net | 81.9* | 1.867 | 58.3* | 2.369 | 61.9* | 2.194 | 3.15 | 0.36 |
| PC-Reg | 82.1* | 0.183 | 58.0* | 0.214 | 61.5* | 0.176 | 6.29 | 0.74 |
| AC-DMiR | 82.3* | 1.739 | 58.4* | 2.174 | 62.2* | 1.863 | 6.88 | 0.79 |
| GL-Net | 82.8* | 1.693 | 58.7* | 1.984 | 62.9* | 1.770 | 5.43 | 0.62 |
| AutoFuse (ours) | 84.5 | 0.026 | 60.2 | **0.025** | 65.0 | 0.026 | 4.82 | 0.59 |
| AutoFuse-Trans (ours) | **84.7** | **0.024** | **60.8** | **0.025** | **65.4** | **0.022** | 7.97 | 0.96 |

**Bold**: the best result in each column. *: $P$<0.05, in comparison to AutoFuse. ↑: the higher is better. ↓: the lower is better.

Table 6. Quantitative comparison for semi-supervised cardiac image registration.

| Method | ACDC (%) | | Runtime (second) | |
|---|---|---|---|---|
| | DSC ↑ | NJD ↓ | CPU ↓ | GPU ↓ |
| Before registration | 59.0* | / | / | / |
| WS-VM | 79.8* | 0.427 | **0.36** | **0.04** |
| RSegNet | 83.0* | 0.089 | 0.45 | 0.05 |
| JPR-Net | 82.6* | 0.442 | 0.52 | 0.06 |
| PC-Reg | 82.5* | 0.117 | 1.36 | 0.14 |
| AC-DMiR | 82.9* | 0.419 | 1.57 | 0.15 |
| GL-Net | 83.2* | 0.385 | 1.24 | 0.13 |
| AutoFuse (ours) | 85.7 | 0.062 | 1.02 | 0.12 |
| AutoFuse-Trans (ours) | **86.2** | **0.047** | 2.21 | 0.23 |

**Bold**: the best result in each column. *: $P$<0.05, in comparison to AutoFuse.
↑: the higher is better. ↓: the lower is better.

### 5.2.2 Ablation Study

Table 7 presents the DSC results of the ablation study for inter-patient brain image registration in the semi-supervised setting. The NJD results are omitted as all methods adopted the same regularization settings and achieved similar NJD results. Compared to employing loss fusion alone, incorporating feature and input fusion both improved the DSC results, where feature fusion contributed to larger DSC improvements than input fusion. The baseline method that employed all loss, feature, and input fusion achieved the highest DSC results among baseline methods. Nevertheless, our AutoFuse (the last row in Table 7) achieved significantly higher DSC results ($P$<0.05) than all the baseline methods. Also, ELK blocks were not used in this ablation study for a fair comparison.

Table 7. DSC results of the ablation study for semi-supervised brain image registration.

| Loss fusion | Feature fusion | Input fusion | Fusion Gate‡ | OASIS (%) | Mindboggle (%) | Buckner (%) |
|---|---|---|---|---|---|---|
| √ | × | × | × | 80.5* | 56.6* | 61.1* |
| √ | √ | × | × | 81.8* | 57.6* | 62.3* |
| √ | × | √ | × | 81.1* | 57.1* | 61.8* |
| √ | √ | √ | × | 82.0* | 58.0* | 62.6* |
| √ | √ | √ | √ | **83.7** | **59.5** | **64.5** |

**Bold**: the highest DSC in each column. *: $P$<0.05, in comparison with AutoFuse. ‡: without ELK blocks.

### 5.2.3 Semi-supervised Learning Analysis

Fig. 5 delineates the DSC results of the AutoFuse trained with different numbers of labeled images for inter-patient brain image registration in the semi-supervised setting. A total of 2656 unlabeled images and 314 labeled images (from OASIS) were used in this analysis, where different numbers of labeled images with or without unlabeled images were used to train our AutoFuse. As shown in Fig. 5, leveraging both labeled and unlabeled images enabled higher DSC results than leveraging unlabeled images alone, in which the DSC results became higher as the number of labeled images increased. In addition, compared to using labeled images alone, leveraging both unlabeled and labeled images also enabled our AutoFuse to achieve better DSC results.

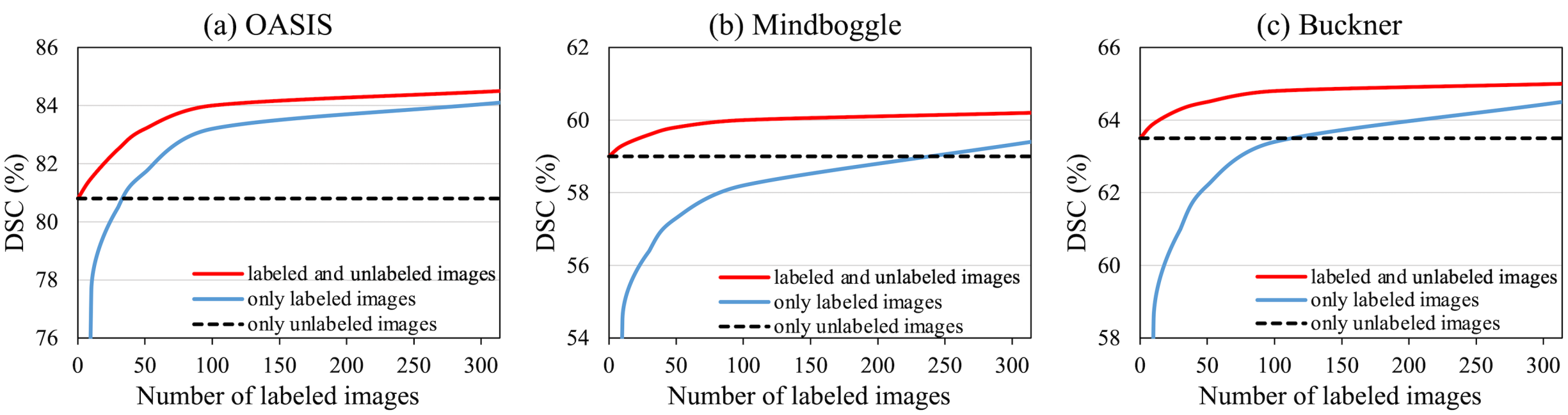

Fig. 5. DSC results of the AutoFuse trained with varying numbers of labeled images together with/without a fixed number (2656) of unlabeled images, for semi-supervised brain image registration on the OASIS (a), Mindboggle (b), and Buckner (c) testing sets.

### 5.3. Qualitative Comparison

Fig. 6 and Fig. 7 present the qualitative comparison between the AutoFuse and existing registration methods for brain and cardiac image registration, in which the AutoFuse-Un and AutoFuse-Semi denote the AutoFuse in the unsupervised and semi-supervised settings. As shown in Fig. 6 and Fig. 7, compared to the existing unsupervised and semi-supervised registration methods, the results produced by our AutoFuse are more consistent with the fixed image, resulting in cleaner error maps for both benchmark tasks.

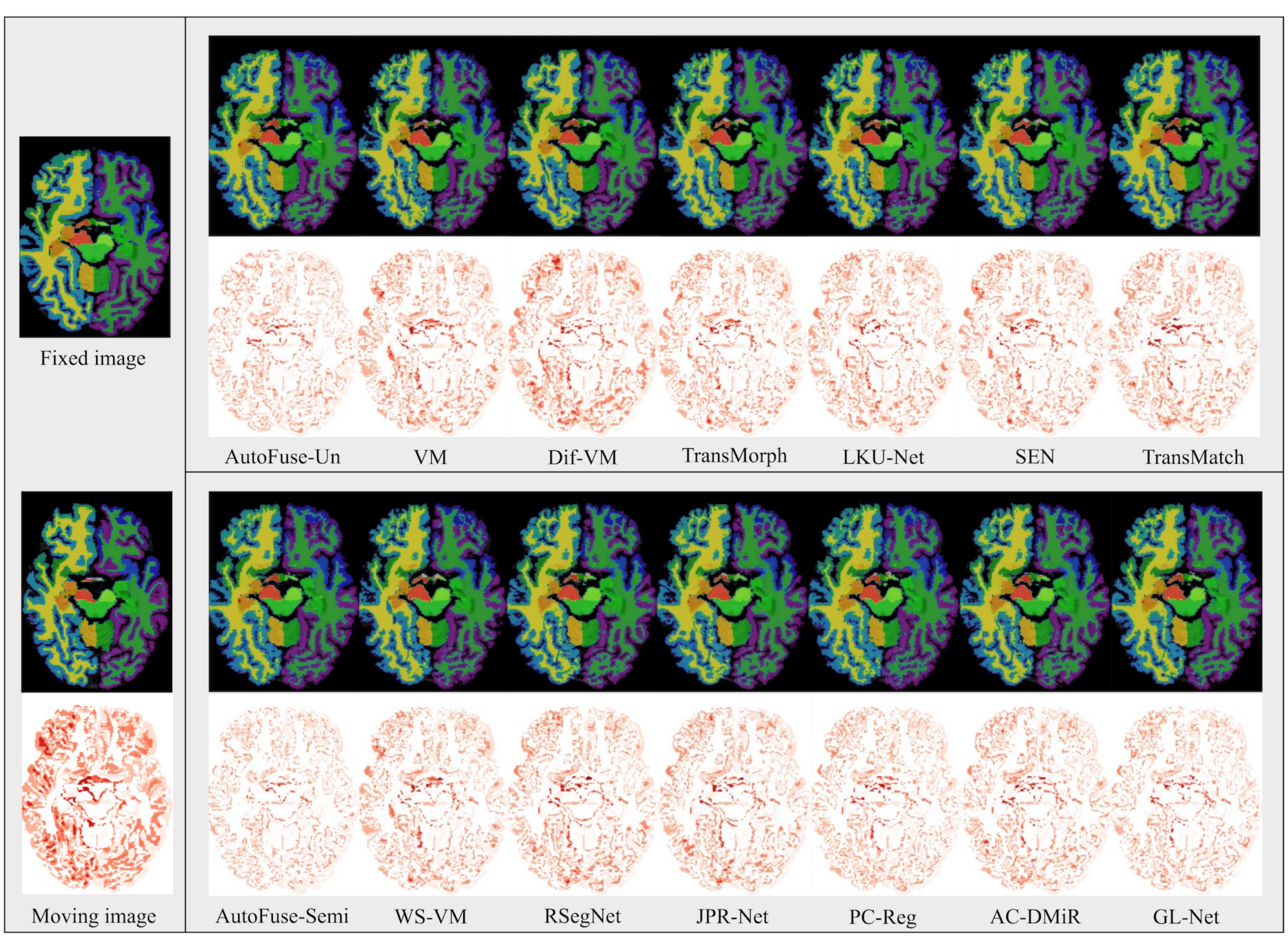

Fig. 6. Qualitative comparison for unsupervised (upper row) and semi-supervised (bottom row) brain image registration. The exemplified image pair is obtained from the Buckner testing set with the labeled anatomical structures in color. Below each image is an error map that shows the differences in segmentation labels between the corresponding image and the fixed image. A cleaner error map indicates better registration.

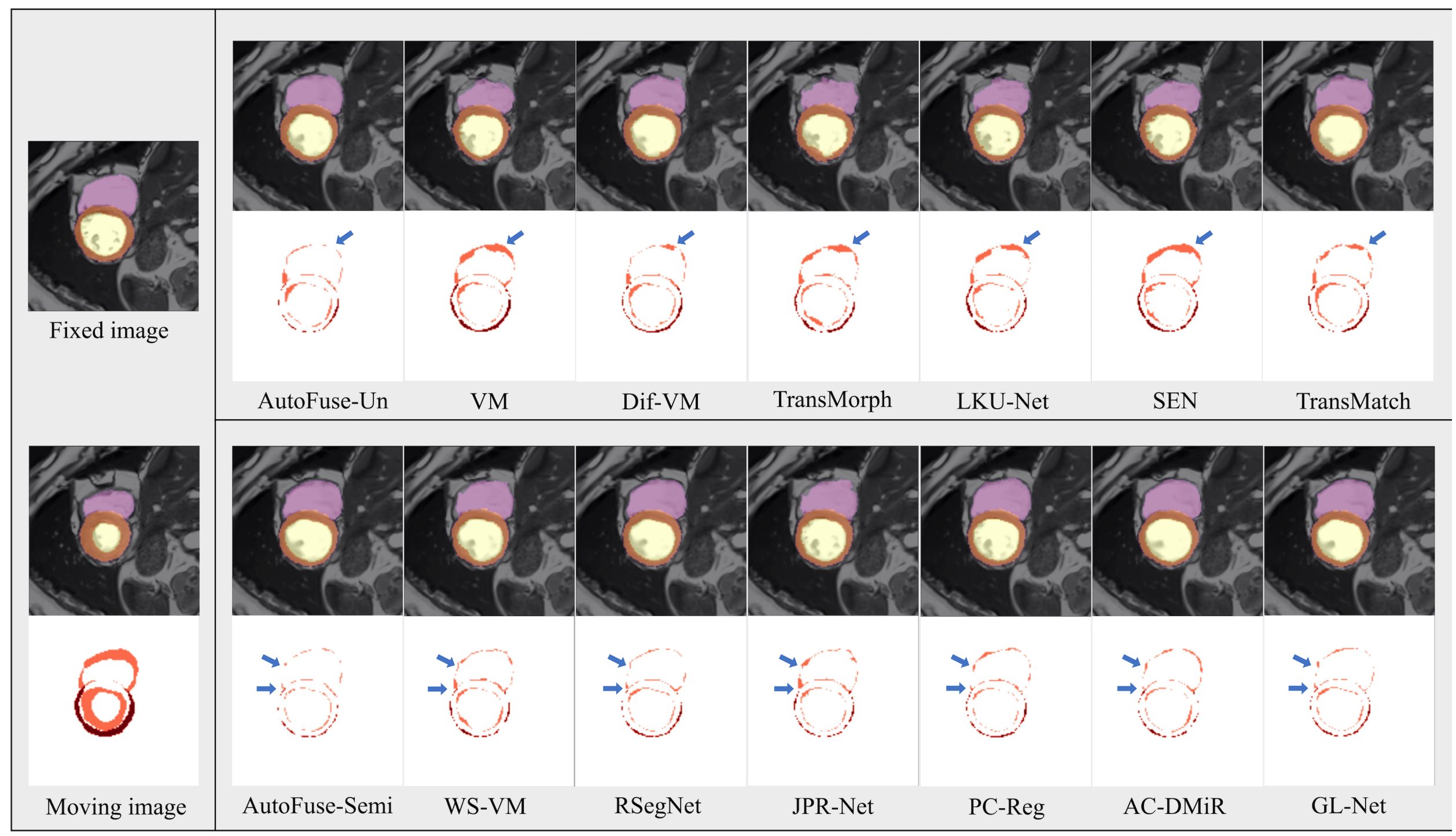

Fig. 7. Qualitative comparison for unsupervised (upper row) and semi-supervised (bottom row) cardiac image registration. The exemplified image pair is obtained from the ACDC testing set with the labeled anatomical regions in color. Below each image is an error map that shows the differences in segmentation labels between the corresponding image and the fixed image. A cleaner map indicates a better registration. The arrows highlight the places where our AutoFuse outperforms the existing registration methods.

### 5.4. Model Interpretation

Table 8 shows the mean values of the adaptive weight maps $w_{fuse}^{i}$ in FG modules, where the mean values of $w_{mf}^{i}$ are omitted as the sum of $w_{fuse}^{i}$ and $w_{mf}^{i}$ is equal to 1. The mean values are 0.5 initially and then are optimized during training, which could be regarded as indicators representing the ratio of information that each FG module fuses from different sources (i.e., $B_{m/f}$ or $B_{fuse}$). For example, a mean value of $w_{fuse}^{i}$ lower than 0.5 indicates that the FG module $\mathcal{F}_i$ tends to fuse less information (<50%) from the $F_{fuse}^{i}$ in $B_{fuse}$ and more information (>50%) from the $B_m$ and $B_f$. As shown in Table 8, our AutoFuse trained with the same training data for brain image registration was evaluated on different testing sets and performed subtly different fusion behaviors for each testing set during inference. In addition, our AutoFuse tended to fuse less information from the $B_{fuse}$ and more information from the $B_m$ and $B_f$ at the FG modules with lower feature resolutions.

Table 8. The mean value of the adaptive weight maps in FG modules during inference on different testing sets.

| Testing set | $w_{fuse}^{2}$ | $w_{fuse}^{3}$ | $w_{fuse}^{4}$ | $w_{fuse}^{5}$ | $w_{fuse}^{6}$ | $w_{fuse}^{7}$ | $w_{fuse}^{8}$ |
|---|---|---|---|---|---|---|---|
| OASIS | 0.546 | 0.384 | 0.212 | 0.163 | 0.187 | 0.322 | 0.507 |
| Mindboggle | 0.559 | 0.375 | 0.205 | 0.184 | 0.206 | 0.331 | 0.518 |
| Buckner | 0.548 | 0.388 | 0.211 | 0.166 | 0.182 | 0.315 | 0.510 |

## 6. Discussion

Our main findings are (i) Our AutoFuse outperformed the state-of-the-art unsupervised and semi-supervised registration methods across two benchmark tasks of brain and cardiac image registration, (ii) Our data-driven fusion strategy outperformed existing empirically-designed fusion strategies in the both unsupervised and semi-supervised settings, (iii) Our AutoFuse can leverage both labeled and unlabeled training data to improve the registration performance through semi-supervised learning, and (iv) Our AutoFuse can learn an adaptive fusion strategy based on training data and showed good generalizability to different unseen testing datasets by self-adapting its fusion strategy during inference.

Both the quantitative and qualitative comparisons on two benchmark tasks demonstrate that our AutoFuse outperformed the state-of-the-art unsupervised and semi-supervised registration methods, including the recent registration methods that employ sophisticated empirically-designed fusion strategies, e.g., TransMatch [34], AC-DMiR [35]. In the quantitative comparison for brain image registration (Table 1 and Table 5), the Mindboggle and Buckner datasets are more difficult than the OASIS dataset as they were used for independent testing and included more anatomical structures for evaluation. This incurred consistently lower DSC results before/after registration on the Mindboggle and Buckner datasets, compared to the OASIS dataset. Moreover, there usually exists a trade-off between DSC and NJD in inter-patient brain image registration [10][14]. Imposing diffeomorphic constraints on spatial transformations can improve smoothness and invertibility, but this unavoidably limits the flexibility of spatial transformations and tends to degrade registration accuracy. Nevertheless, our AutoFuse, as a diffeomorphic registration method, achieved both the best DSC and NJD results, which demonstrates that our AutoFuse can perform accurate registration with highly smooth and invertible spatial transformations. In the quantitative comparison for cardiac image registration (Table 2 and Table 6), diffeomorphic registration methods showed advantages as intra-patient cardiac images carry invertible deformations without topology corruption. As a diffeomorphic method, our AutoFuse achieved better registration performance than all the comparison methods, including the compared diffeomorphic registration methods (DifVM and DTN). In addition, we found that leveraging transformer blocks in the AutoFuse produced consistent improvements for both unsupervised and semi-supervised registration on both two benchmark tasks, which suggests that our data-driven fusion strategy is generalizable to both CNN and transformer architectures.

In the ablation study for unsupervised registration (Table 3), the LateFuse achieved the worst DSC results, which suggests that late fusion is insufficient to characterize the complex spatial correspondence between images. The EarlyFuse and MidFuse fuse information at earlier stages and thus achieved better DSC results than the LateFuse. The MultiFuse fused information at all scales, which enabled it to achieve better DSC results than other baseline methods. Nevertheless, the MultiFuse is still limited by an empirically-designed fusion strategy that restricts the information fusion to manually-defined prior knowledge. Our AutoFuse leveraged FG modules to realize a data-driven fusion and thus achieved the best DSC results among all the methods. It should be noted that, for a fair comparison, we purposely adjusted each baseline method, so that they have similar or higher parameter numbers than AutoFuse (refer to Appendix C for detailed architectural settings). Moreover, ELK blocks were not used in this ablation study. These results suggest that the improvements of AutoFuse did not result from the use of extra parameters or ELK blocks, and our data-driven fusion strategy contributed to these improvements while not adding extra parameters.

In the ablation study for semi-supervised registration (Table 7), our AutoFuse also achieved better DSC results than all the baseline methods that employ the existing loss, feature, and input fusion strategies. ELK blocks were not used in this ablation study for a fair comparison and all the baseline methods share the same three-branch encoder-decoder architecture as the AutoFuse (detailed in Appendix C). These results further suggest that the improvements of AutoFuse indeed resulted from the use of our data-driven fusion strategy, and our data-driven fusion strategy outperformed the existing loss, feature, and input fusion strategies and their combinations in the semi-supervised setting. We also found that, compared to the unsupervised setting, our AutoFuse achieved significantly higher DSC results in the semi-supervised setting, which demonstrates that our AutoFuse can effectively leverage the anatomical information in segmentation labels to improve registration. The semi-supervised learning analysis (Fig. 5) also validates this finding, which shows that leveraging both labeled and unlabeled training data enabled our AutoFuse to achieve better registration performance than using labeled or unlabeled training data alone.

In addition, the quantitative comparison for brain image registration (Table 1 and Table 5) shows that our AutoFuse was well-generalized to two independent testing sets (Buckner and Mindboggle) and produced consistent improvements. This is attributed to the fact that our AutoFuse can learn an adaptive fusion strategy. As shown in the model interpretation (Table 8), the AutoFuse adapted its fusion strategy for different testing sets during inference. Furthermore, we identified that the AutoFuse-learned fusion strategy can provide insights to facilitate the understanding of fusion strategy design. For example, as shown in Table 8, the trained AutoFuse tended to fuse less information from the fusion branch ($B_{fuse}$) and grab more information from the two feature extraction branches ($B_m$ and $B_f$) at the FG modules with lower feature resolutions. This reveals an insight that the afore-fused inter-image spatial information (in the fusion branch) has deteriorated after downsampling operations and, therefore, the unfused intra-image spatial information (in the two feature extraction branches) is needed to rebuild the spatial correspondence between images. This insight is consistent with the recent sophisticated fusion strategies [34][36], where the spatial information of each image was extracted via separate encoders and then fused at multiple scales to explore their spatial correspondence. Nevertheless, compared to these empirically-designed fusion strategies, our data-driven fusion strategy provides more flexibility to optimize the fusion strategy based on data, where the information can be fused using the learned weight maps at each scale.

Our study has a few limitations. First, by using a data-driven fusion strategy, we expect that our AutoFuse can search over a large search space to find the optimal fusion strategy based on training data. However, despite the current search space being vast and including commonly-adopted fusion strategies, this study mainly focused on the optimization of fusion location (i.e., where to fuse the information within the network?), while the fusion operations were not fully investigated. In our future study, we will investigate further inclusion of other fusion operations, e.g., cross-attention transformers [15][34], into the search space. Moreover, this study mainly focused on brain and cardiac image registration. Our AutoFuse can be further validated with other registration applications, e.g., intra-patient lung registration and glioma registration [61].

## 7. Conclusion

In this study, we have outlined a data-driven fusion strategy in an Automatic Fusion Network (AutoFuse) for deformable image registration. Unlike existing deep registration methods that adopt empirically-designed fusion strategies, our AutoFuse employs Fusion Gate (FG) modules to control the information fusion, which optimizes its fusion strategy based on training data for both unsupervised and semi-supervised registration. Extensive experiments on two well-benchmarked medical registration tasks (inter-patient brain image registration and intra-patient cardiac image registration) show that our AutoFuse outperforms state-of-the-art unsupervised and semi-supervised registration methods on both registration accuracy and transformation invertibility.

## Acknowledgements

This work was supported by Australian Research Council (ARC) under Grant DP200103748. The computations in this paper were run on the π 2.0 cluster supported by the Center for High Performance Computing at Shanghai Jiao Tong University.

## Author contributions: CRediT

Mingyuan Meng: Conceptualization, Formal analysis, Investigation, Methodology, Visualization, Writing – original draft

Michael Fulham: Validation, Writing – review and editing

Dagan Feng: Conceptualization, Supervision

Lei Bi: Conceptualization, Project administration, Writing – review and editing

Jinman Kim: Funding acquisition, Project administration, Supervision, Writing – review and editing

## Appendix A: Architecture Details

Table A1 presents the kernel numbers of AutoFuse used in the experiments. In the ablation studies, ELK blocks were not used while other kernel numbers were unchanged. In addition, Table A2 presents the architectural setting of AutoFuse-Trans used in the experiments, including embedding dimensions, attention head numbers, and window size. These settings were empirically chosen to fit into our current GPU memory, which could be adjusted to fit into other computing devices.

Table A1. Kernel numbers of the AutoFuse used in the experiments.

| $i =$ | 1 | 2 | 3 | 4 | 5 | 6 | 7 | 8 | 9 |
|---|---|---|---|---|---|---|---|---|---|
| $F_m^i$ / $F_f^i$ | 16 | 32 | 32 | 64 | 64 | 64 | 32 | 32 | / |
| $F_{fuse}^i$ | 32 | 64 | 64 | 128 | 128 | 128 | 64 | 64 | 64 |
| $F_{\mathcal{F}}^i$ | / | 64 | 64 | 128 | 128 | 128 | 64 | 64 | / |
| ELK | / | 32 | 32 | 64 | 64 | 64 | 32 | 32 | / |

The last row (ELK) presents the kernel numbers of four parallel convolutional layers in each ELK block.

Table A2. Architecture details of the AutoFuse-Trans used in the experiments.

| Embedding dimensions of $B_{fuse}$ | Embedding dimensions of $B_{m/f}$ | Head numbers | Window size |
|---|---|---|---|
| [64, 128, 256, 512, 256, 128, 64] | [32, 64, 128, 256, 128, 64, 32] | [4, 8, 16, 32, 16, 8, 4] | [5, 5, 5] |

## Appendix B: Segmentation Performance

Fig. A1 presents the segmentation masks predicted by our AutoFuse in the semi-supervised setting, in which the predicted segmentation masks are highly consistent with the ground-truth segmentation labels. Quantitively, our AutoFuse achieved DSC of 0.886 and 0.863 for segmentation on the OASIS and ACDC testing sets. These segmentation results imply that our AutoFuse can well identify the anatomical information in images and leverage this information to improve registration.

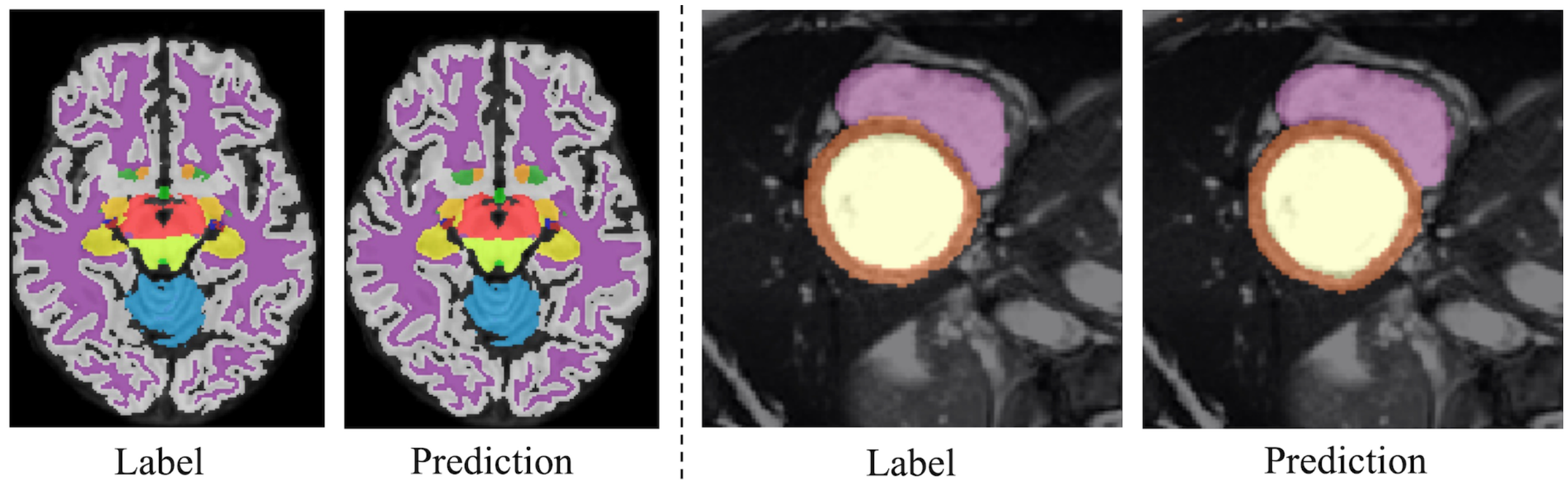

Fig. A1. Examples of the segmentation masks predicted by our AutoFuse in the semi-supervised setting. The ground-truth labels and the predicted masks on the OASIS (left) and ACDC (right) testing sets are presented with the labeled anatomical structures/regions colored.

## Appendix C: Baseline Methods in Ablation Studies

Fig. A2 illustrates the architecture of the baseline methods used in the ablation studies, including eight baseline methods in the unsupervised and semi-supervised settings. The kernel numbers of each baseline method are also shown in Fig. A2.

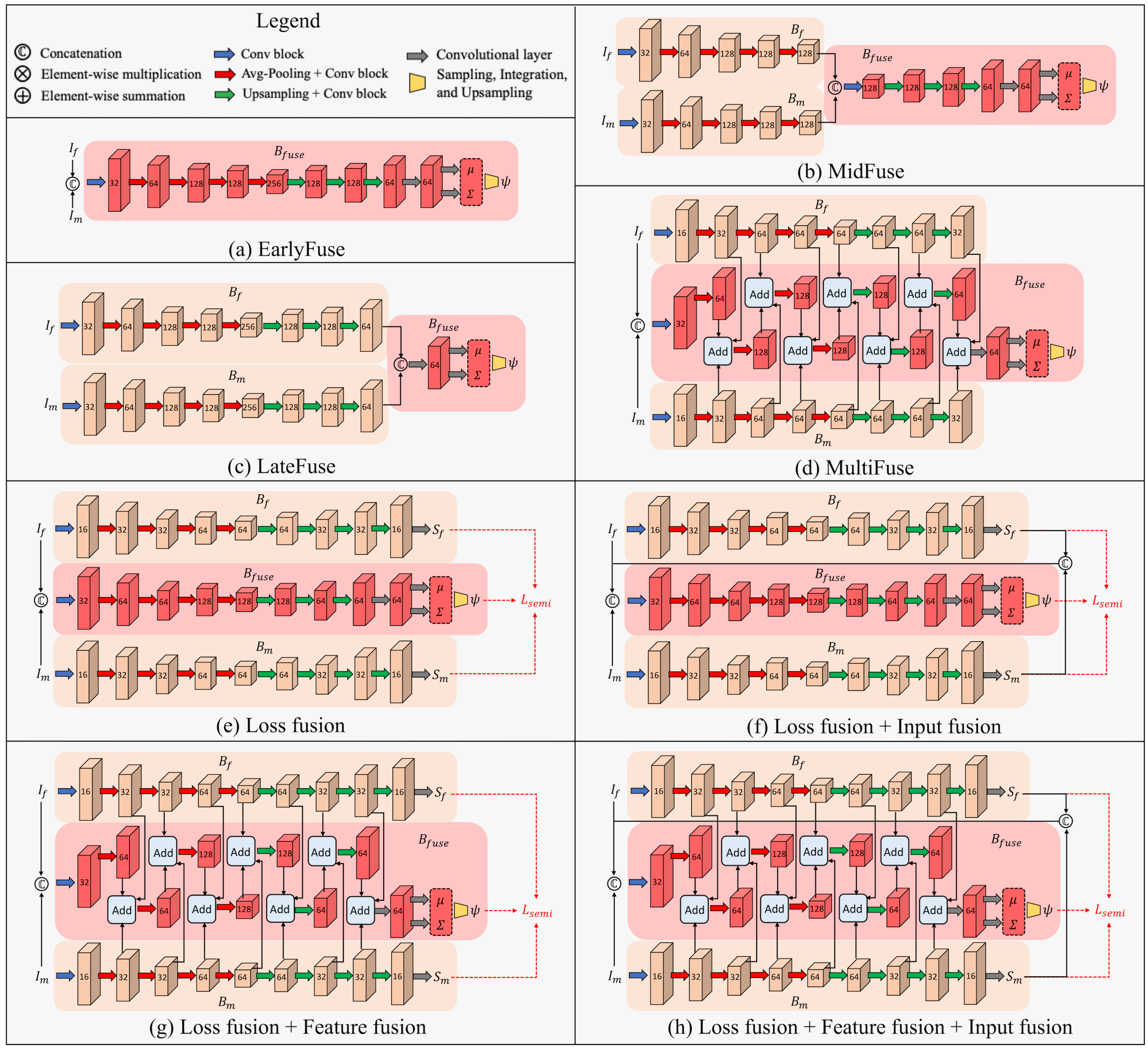

Fig. A2. Architectures of the baseline methods in the ablation studies. Each cube represents the feature maps generated from its preceding convolutional layer with the kernel numbers shown within the corresponding cube. The skip connections of each branch are omitted in this figure for the sake of the clarity. The branches $B_m$ and $B_f$ share the same weights.